# Quantifying Capacity Loss due to Solid-Electrolyte-Interphase Layer Formation on Silicon Negative Electrodes in Lithium-ion Batteries

Siva P. V. Nadimpalli,[a] Vijay A. Sethuraman,[a] Swapnil Dalavi,[b] Brett Lucht,[b] Michael J. Chon,[a] Vivek B. Shenoy,[a] and Pradeep R. Guduru[a],*

[a]School of Engineering, Brown University, 182 Hope Street, Providence, Rhode Island 02912
[b]Department of Chemistry, The University of Rhode Island, Kingston, Rhode Island 02881

*Corresponding author: Pradeep R. Guduru, E-mail: Pradeep_Guduru@brown.edu, Telephone: +1 401 863 3362, Fax: +1 401 863 9009

**Abstract**

Charge lost per unit surface area of a silicon electrode due to the formation of solid-electrolyte-interphase (SEI) layer during initial lithiation was quantified, and the species that constitute this layer were identified. Coin cells made with Si thin-film electrodes were subjected to a combination of galvanostatic and potentiostatic lithiation and delithiation cycles to accurately measure the capacity lost to SEI-layer formation. While the planar geometry of amorphous thin films allows accurate calculation of surface area, creation of additional surface by cracking was prevented by minimizing the thickness of the Si film. The cycled electrodes were analyzed with X-ray photoelectron spectroscopy to characterize the composition of the SEI layer. The charge lost due to SEI formation measured from coin cell experiments was found to be in good agreement with the first-cycle capacity loss during the initial lithiation of a Si (100) crystal with planar geometry. The methodology presented in this work is expected to provide a useful practical tool for battery-material developers in estimating the expected capacity loss due to first cycle SEI-layer formation and in choosing an appropriate particle size distribution that balances mechanical integrity and the first cycle capacity loss in large volume expansion electrodes for lithium-ion batteries.

## 1 Introduction

Silicon is considered to be a promising material as a negative electrode for the next generation high-energy-density lithium-ion batteries due to its high capacity (*ca.* 3579 mAh/g) [1]. However, large volume changes during lithiation and delithiation are known to cause fracture and capacity loss, which has been an obstacle in its wider use in commercial Li-ion battery applications [2]. For example, studies on electrodes made of Si bulk films and particles of the order of microns revealed severe capacity fade and short cycle life due to pulverization [3,4]. On the other hand, electrodes made of Si thin-films, nano-wires, and nano-particles showed a marked improvement in the fracture performance [5,6]. Clearly, the surface area per unit mass increases in inverse proportion to the particle size. Although decreasing the particle size improves the rate of lithiation/delithiation and fracture resistance, it also offers large surface area for electrolyte-reduction reactions resulting in the formation of solid-electrolyte-interphase (SEI) layer and the associated irreversible loss of Li. Dahn *et al.* [7] showed that capacity loss to SEI formation on graphite electrodes was proportional to the surface area of the electrode; assuming the formation of a $Li_2CO_3$ film, they calculated an SEI thickness of 4.5 nm on carbon particles, consistent with the barrier thickness needed to prevent electron tunneling.

SEI layer plays an important role in the safety, power capability, and cyclic life of Li-ion batteries [8-12]. In one of the earliest works on the SEI, Peled [11] proposed a model for SEI formation mechanism in non-aqueous electrochemical systems such as Li-ion batteries and concluded that formation of a chemically and mechanically stable SEI layer is the key for improving the cycle life of batteries. For example, Chen *et al.* [13] enhanced the electrochemical performance of Si electrodes by improving the properties of SEI layer (achieved by adding vinylene carbonate additive in their electrolyte). Recently, it was shown that additives such as propylene carbonate, lithium difluoro-oxalatoborate, and fluoro-ethylene carbonate dramatically improve the cyclic efficiency of Si electrodes [14-17]. Lee *et al.* [12] found that SEI layer on Si electrode forms due to reduction of organic solvents and anions at the electrode surface during charging and discharging cycles of batteries; bulk of the formation occurs during the first cycle. Yu-Chan *et al.* [18] characterized SEI layers formed on Si electrodes and found fluorinated C and Si species, besides the usual $Li_2CO_3$, alkyl Li carbonates ($ROCO_2Li$), LiF, ROLi, and polyethylene oxides that are found on graphite electrodes. SEI formation on the negative electrode is an irreversible reaction that consumes cyclable Li-ions from the positive electrode leading to most of the capacity loss observed in the first lithiation/delithiation cycle of secondary lithium-ion batteries. Besides capacity loss in the first cycle, continuous formation of this layer also increases resistance to Li-ion diffusion (*i.e.*, internal impedance of a battery) [12].

In spite of the important role played by the SEI layer on the calendar and cycle life of secondary lithium-ion batteries made with Si anodes, there have not been many studies on understanding the mechanisms of initial formation of SEI on Si electrodes. Furthermore, there have been few attempts towards quantifying the first-cycle capacity loss due to SEI-layer formation on Si. Since the requirements for fracture tolerance (*i.e.,* small particle size) would be in conflict with the need to minimize the first cycle irreversible loss (*i.e.,* minimize the surface area per unit mass), measurement of Li loss per unit area due to irreversible reactions can serve as a useful design parameter in arriving at optimal micro/nano architectures for Si-based electrodes. Furthermore, quantifying the charge lost to SEI formation is essential for accurately arriving at the true state-of-charge of the silicon electrode during its initial lithiation. For *in situ* measurements of stress as well as the mechanical properties of a silicon electrode [19-21] during its initial lithiation, it is essential to know the exact Li concentration in Si electrode to calculate its volumetric strain.

The objective of this study is to devise a method to quantify the charge loss per unit area of Si electrodes during the first half-cycle due to the formation of the SEI-layer and to identify the species that constitute this layer. Coin cells were made with Si thin-film as working electrode and Li metal as counter/reference electrodes, and they were cycled to measure the first cycle capacity loss. The planar geometry of thin films allowed accurate calculation of surface area; further, creation of additional surface by cracking was prevented by minimizing the thickness of the Si film. The cycled electrodes were analyzed with X-ray photo electron spectroscopy to characterize the species decomposed on Si electrode. The measured first cycle capacity loss from coin cell experiments was found to be good agreement with that obtained from the initial lithiation of a Si (100) crystal with planar geometry.

## 2 Experiments

### 2.1. Si electrode fabrication and electrochemical experiments

Electrodes were fabricated by depositing *ca*. 200 nm thick Cu layer (current collector) on a (1 cm x 1 cm, 500 μm nominally thick, and double side polished) Si substrate with *ca*. 500 nm thick thermally grown oxide on all sides [shown in Fig. 1 (a)], followed by a 20 nm thick Si film, which was deposited by RF-magnetron sputtering at 150W power and less than 2 mTorr Ar pressure.

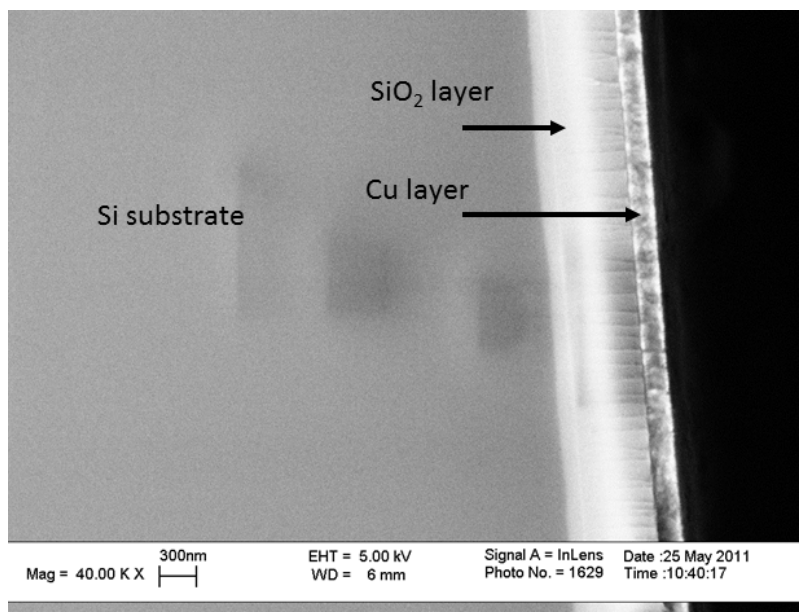

*Figure 1(a): SEM image of electrode cross section showing Si substrate with ca. 500 nm SiO$_2$ and ca. 200 nm thick Cu layer.*

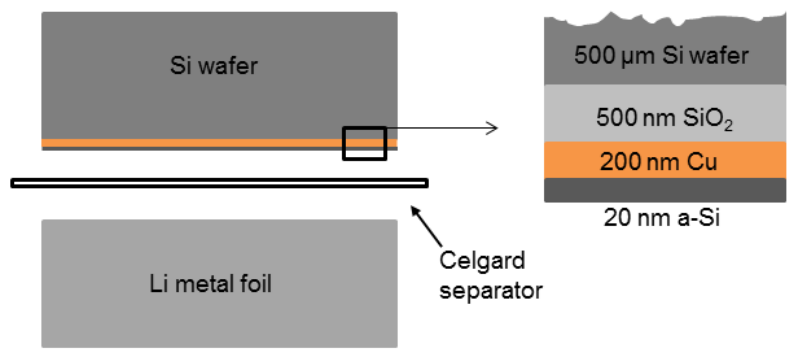

*Figure 1(b): Schematic illustration of the coin cell made of Si thin film electrode and Li metal counter electrode; note that the schematic is not drawn to scale.*

The thermally grown oxide on the Si substrate acts as a barrier for Li diffusion, isolating the substrate from participating in the electrochemical reactions. The polished Si substrate provides a flat and smooth surface so that the surface roughness can be neglected while calculating the surface area of the sputtered Si film. The sputtered Si films under these conditions are known to be amorphous [1]. To confirm this, Raman spectra were obtained on the magnetron-sputtered Si thin-films using a LabRAM Scope (Horiba Scientific) with a HeNe laser (λ = 632.8 nm, 1 mW power) as the excitation source and are compared to that obtained from single-crystal Si wafers in Fig. 1(c); absence of any peaks from the sputtered Si film clearly indicates that it is amorphous in nature. Coin cells were fabricated in which the Si film was the working electrode and a Li metal foil served as the counter/reference electrode.

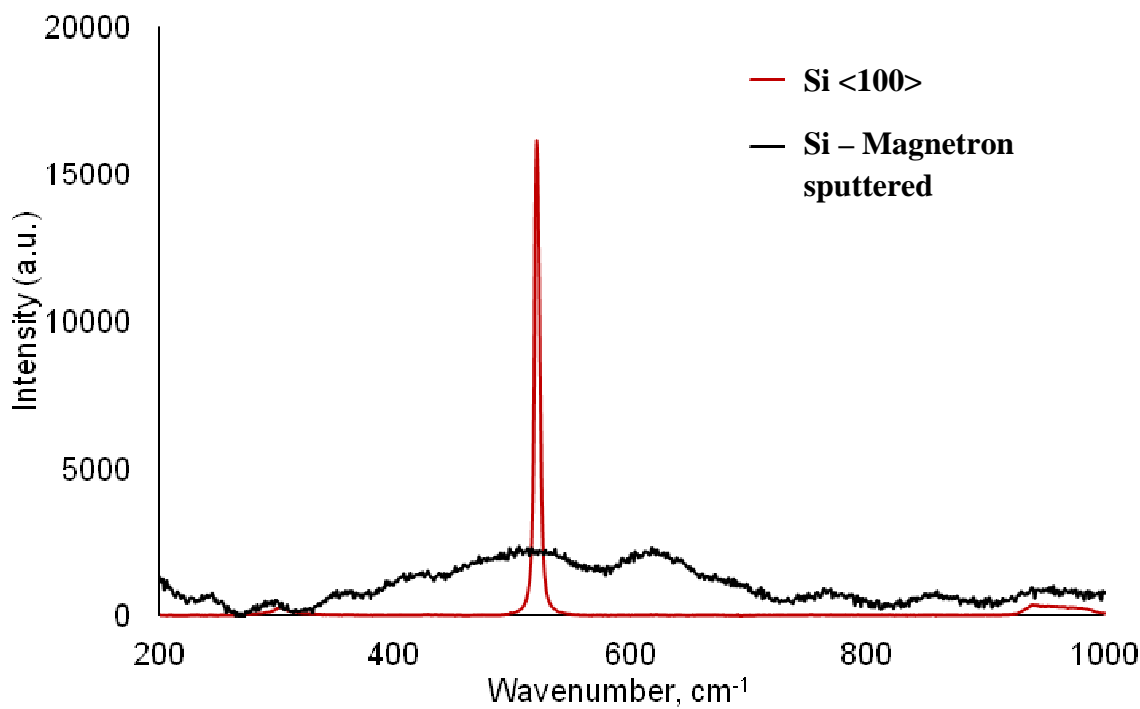

*Figure 1(c): Comparison of Raman spectra of magnetron-sputtered Si thin-films and boron-doped <100> Si wafer showing the former is amorphous.*

The electrolyte was a solution of 1M $LiPF_6$ in ethylene carbonate (EC):diethyl carbonate (DEC):dimethyl carbonate(DMC) (1:1:1 by weight). The cells were assembled in an argon-filled glove box maintained at 25°C; sample geometry is schematically illustrated in Fig. 1(b). The assembled coin cells were cycled (*i.e.*, lithiation and delithiation) using a Solartron 1470E Multistat at 25 °C (±1°C) to measure the capacity loss of the cell during the first cycle. The following two types of experiments were conducted:

(a) The Si electrode was lithiated galvanostatically at 5μA $cm^{-2}$ until the potential decreased to pre-determined cut-off levels, which were chosen to be 0.3 V, 0.2 V, and 0.15 V *vs.* Li/$Li^+$ in three different samples. This was followed by a five minute open-circuit-potential relaxation, and delithiation at a constant current density of 5μA/$cm^2$ until the potential increased to 1.2 V *vs.* Li/$Li^+$; samples were then held at 1.2 V until the current became very low (*i.e.*, less than 0.005 μA $cm^{-2}$). This procedure ensures almost complete removal of Li from the Si electrode. The potential *vs.* time plot in a representative experiment (with a cut-off potential of 0.2 V *vs.* Li/$Li^+$) is shown in Fig. 2 as the dotted line.

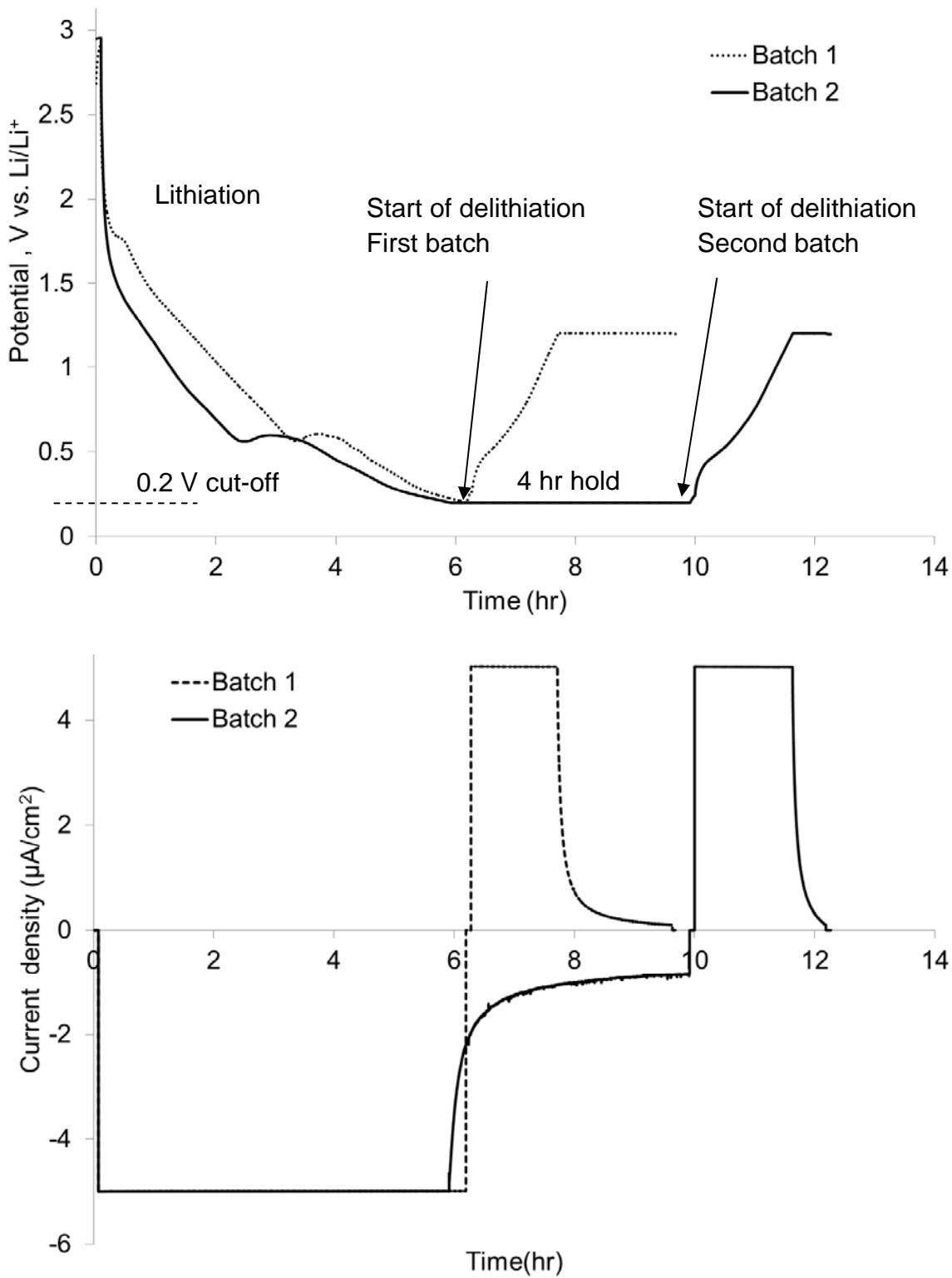

*Figure 2: Examples of electrochemical data of coin cells from each of the two batches showing (a) transient potential response and (b) current density as a function of time. Note that during the 4hr potential hold the current density of cell from Batch 2 decays to 1 µA cm$^{-2}$.*

(b) In the second type of experiment, the cut-off potential for lithiation was fixed at 0.2 V *vs.* Li/Li$^+$ in all samples; however, the hold period at the cut-off potential for different samples was varied between 0 and 10 hr. The delithiation procedure and current densities were the same as the previous experiment. The solid curves in Fig. 2 show an example where the hold time at the cut-off potential was 4 hr.

These cut-off potentials were chosen such that they are not only favorable for SEI formation (or electrolyte reduction reactions) but also suitable for amorphous Si film lithiation. In order to identify the potential in the neighborhood of which the electrolyte reduction reactions begin to occur (leading to the formation of the SEI layer), cyclic voltammogram measurements were conducted. Coin cells with Cu disks (1.5 cm diameter and 3 mm thickness) as working and Li metal foils as counter/reference electrodes were subjected to potential sweep between 0.05 V and 3 V *vs.* Li/Li$^+$ at a scan-rate of 0.5 mV/s and the resulting current response is shown in Fig. 3. Since Cu does not alloy with Li, it was assumed that any sharp increase in the cell current is mainly due to the reduction reaction of the electrolyte. The equilibrium potential below which these reduction reactions take place was identified as the potential at which the current begins to increase rapidly, as indicated by the arrow in Fig. 3. The equilibrium potential was determined to be approximately 0.8 V *vs.* Li/Li$^+$, which was well above the range used in the coin cell experiments (0.15 to 0.3 V *vs.* Li/Li$^+$). We note that this value is consistent with the values reported in the literature [22].

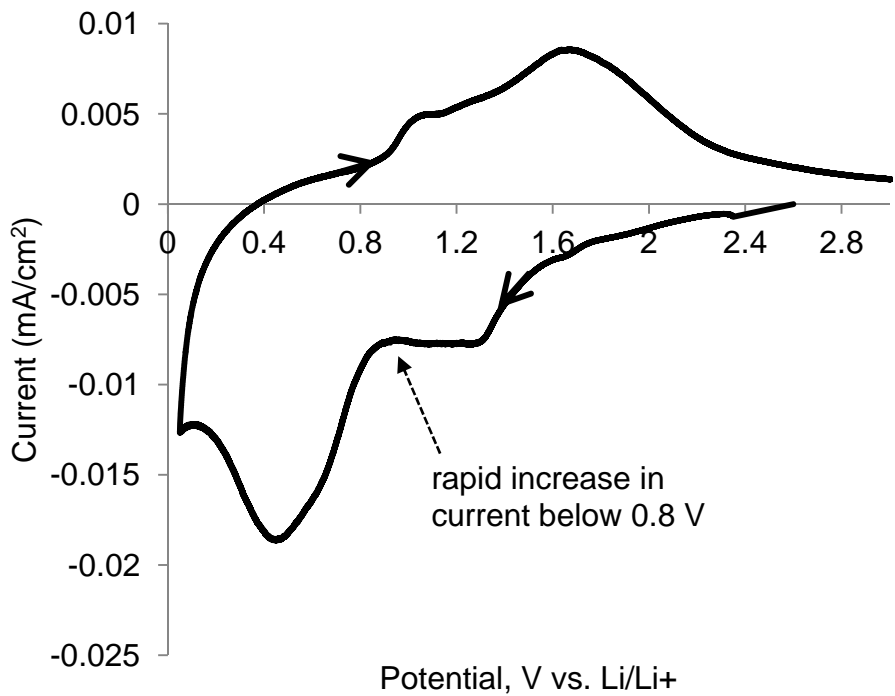

*Figure 3: Cyclic voltammogram of a coin cell with Cu as working electrode, Li metal foil as counter/reference electrode, and 1M of LiPF$_6$ in EC:DEC:DMC (1:1:1) as electrolyte. The voltage was swept (direction of sweep indicated by arrows) between 3 V and 0.05 vs. Li/Li$^+$ at a rate of 0.5 mV/s. Note that the current increases significantly below 0.8 V vs. Li/Li$^+$, indicating that the electrolyte reaction (SEI formation) is dominant below this potential.*

The capacity loss per unit surface area in each experiment was taken to be the difference between the charge supplied to the cell during lithiation and the charge recovered during delithiation, divided by the surface area of the Si electrode,

Q_lost = (Q supplied to the cell - Q recovered from the cell)/(surface area of the electrode )     (1)

Since the surface roughness ($R_a$) of the deposited Si films was extremely small (*ca.* 3 nm as measured by white light interferometry), the real surface area of the film was taken to be approximately equal to its nominal surface area. Further, the Si film thickness of 20 nm was chosen to ensure that the film does not crack during delithiation, so that there is no additional surface area due to cracking on which an SEI layer gets deposited. The integrity of the Si film, *i.e.*, the absence of cracking, after lithiation-delithiation cycle was confirmed by SEM imaging (Fig. 4).

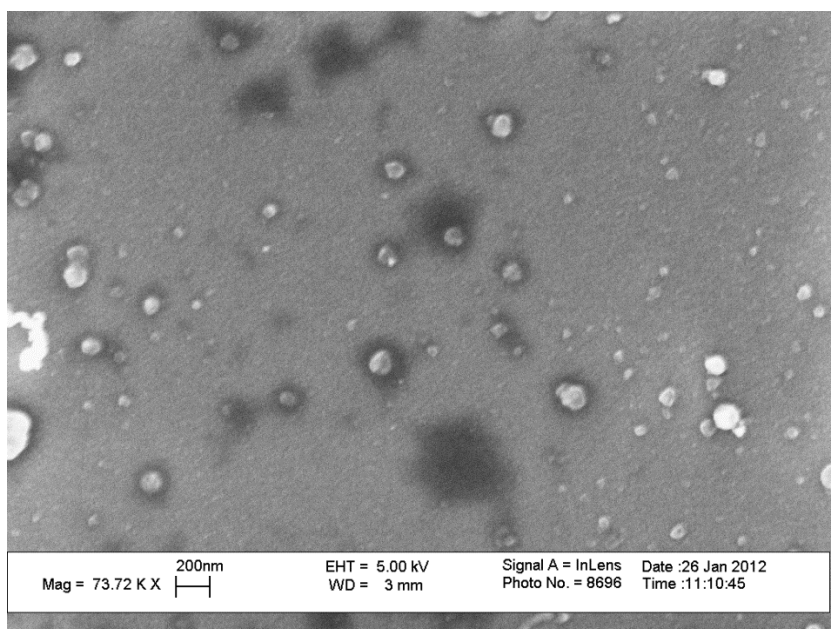

*Figure 4: SEM image of a Si electrode film after a single lithiation and delithiation cycle, showing an intact film with no cracks. The white spots are particles of $LiPF_6$ salt residue.*

In order to verify the estimated $Q_{lost}$ due to SEI (based on measurements from the coin cells), galvanostatic lithiation/delithiation experiments were conducted on Si(100) single-crystal electrodes [polished, 50.8 mm diameter, 500 μm thickness, (100) orientation]. For this, one lithiation/delithiation cycle was carried out galvanostatically on the polished surface of a Si(100) single-crystal electrode; the ratio of delithiation capacity to lithiation capacity was calculated and the charge lost during this first cycle was compared to the above measurements for charge lost due to the SEI formation. For experimental details on the lithiation and delithiation of single-crystal Si electrodes, please see Ref.[23].

## 2.2 Surface characterization of Si electrodes

Following cycling, the coin cells were disassembled in Argon atmosphere for surface characterization. The electrodes were gently rinsed with anhydrous di-methyl carbonate (DMC) to remove residual carbonate solvents and $LiPF_6$ salt, followed by vacuum drying overnight at room temperature. X-ray photo electron spectroscopy (XPS) analysis was conducted with a PHI 5500 system using Al Kα (hυ=1486.6 eV) radiation under ultrahigh vacuum. All peaks were referenced to universal carbon contamination peak at 285 eV. The spectra obtained were analyzed by Multipak 6.1A software. Line syntheses of elemental spectra were conducted using Gaussian–Lorentzian (70:30) curve fit with Shirley background subtraction. Element concentration was calculated based on the equation $C_x = (I_x/S_x)/(\Sigma I_i/S_i)$, where $I_x$ is the intensity of the relative element, and $S_i$ is the sensitivity number of the element, acquired from the Multipak 6.1A software.

## 3 Results and Discussion

### 3.1 Capacity loss to SEI layer formation

$Q_{lost}$ from the first type of coin-cell experiments is plotted against the cut-off potential in Fig.5, and the results from the second type of experiments are plotted in Fig. 6 in which the *x*-axis is the hold time at the cut-off potential of 0.2 V *vs.* Li/Li$^+$. Each data point (filled square) in these plots represents a single coin-cell experiment. Figure 5 shows that the charge lost per unit area due to SEI layer varied between 0.064 and 0.1 C cm$^{-2}$ (i.e., 0.0178 and 0.0278 mAh cm$^{-2}$), with a mean value of *ca.* 0.082 C cm$^{-2}$ (0.023 mAh cm$^{-2}$).

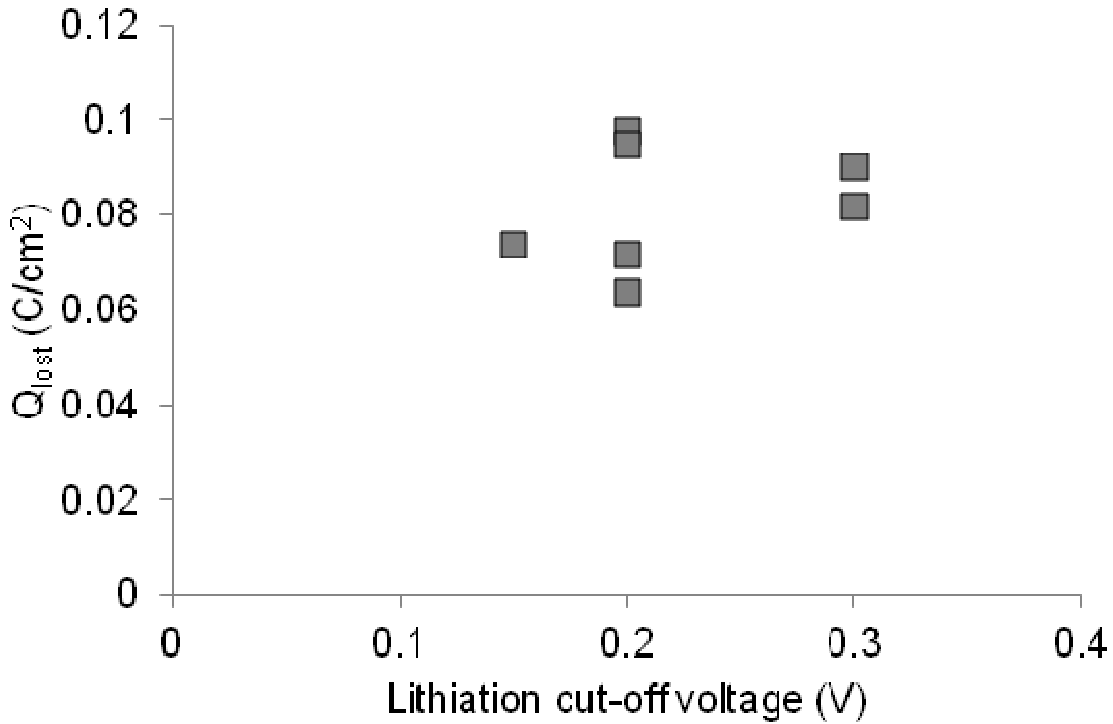

*Figure 5: Loss of charge as a function of different cut-off potential levels (data from first batch of experiments). Note: 1 C/cm$^2$ = 0.278 mAh cm$^{-2}$.*

Results from the second batch of experiments (Fig.6) show that $Q_{lost}$ increased slightly with the hold time for about 4 hr, beyond which it reached a plateau of about 0.09 C cm$^{-2}$ (0.025 mAh cm$^{-2}$).

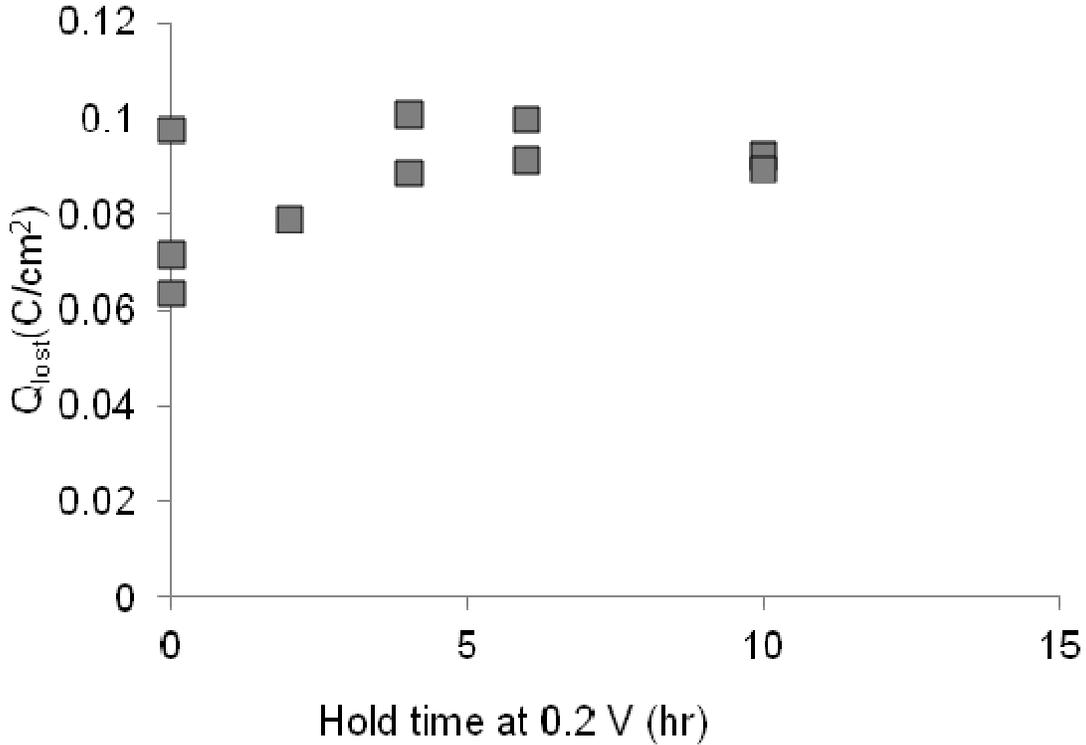

*Figure 6: Loss of charge as a function of hold time in the second batch of coin cell experiments. Note: 1 C cm$^{-2}$ = 0.278 mAh cm$^{-2}$.*

The average $Q_{lost}$ for the polished Si wafer samples (an example potential vs. capacity plot at constant current density is shown in Fig. 7) was 0.072 C cm$^{-2}$ (± 0.01 C cm$^{-2}$ of standard deviation, 3 samples) which agrees very well with the range of values in Figs. 5 and 6 and provides additional verification of the $Q_{lost}$ measurements. It should be noted that, although the lithiation mechanisms of 20 nm amorphous Si film and single crystal Si wafer are very different, the charge lost to SEI formation in these two sets of samples was in close agreement.

The main assumption in interpreting the experimental data presented above is that the unrecovered charge in the first lithiation-delithiation cycle is due to the formation of the SEI layer only. This is a reasonable assumption since the film was uncracked and damage-free. However, it is possible that some of the Li is irreversibly bonded in amorphous Si and is not recoverable electrochemically, which introduces an error into the reported results. In view of such uncertainties, the $Q_{lost}$ values reported here should be viewed as an upper bound estimate of Li$^+$ lost due to SEI layer. Further, the $Q_{lost}$ values are known to be sensitive to the electrolyte composition; hence the reported results should be regarded as baseline data for generic Li ion battery electrolytes that do not contain any special additives to control or influence the SEI layer formation.

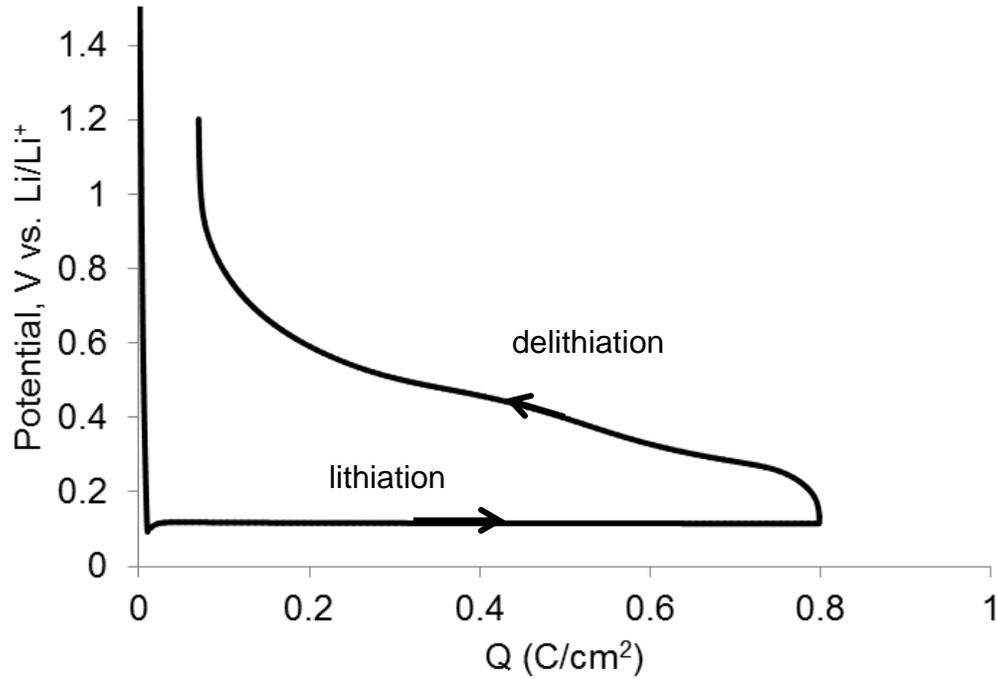

*Figure 7: Potential vs. capacity data corresponding to the initial lithiation and delithiation of Si (100) single crystal electrode at a constant current density of 12.5 µA cm$^{-2}$. The first-cycle efficiency (i.e., ratio of delithiation capacity to lithiation capacity) was 91.2%. Note: 1 C cm$^{-2}$ = 0.278 mAh cm$^{-2}$.*

The main conclusion from the experimental study reported here is that the charge lost due to SEI formation in Li-ion batteries made with Si based negative-electrodes and 1M LiPF$_6$ in (1:1:1) EC: DEC: DMC electrolyte is in the range of 0.06 – 0.1 C cm$^{-2}$ (0.0178 and 0.0278 mAh cm$^{-2}$), which can be used to predict or estimate the first cycle capacity loss for Si electrodes in any other configuration for which the real surface area is known. For example, Fig.8 shows the estimated first-cycle capacity loss in Si particle and wire geometries as a function of particle size, assuming a $Q_{lost}$ value of 0.064 C cm$^{-2}$. Note that the capacity loss is almost negligible for particles with sizes more than 10 µm and increases sharply as the size decreases below 1 µm.

Chan *et al.* [5] reported electrochemical cycling experiments on Si nano-wires and reported the first-cycle capacity loss to be 1153 mAh g$^{-1}$. Taking a value of 205 nm for the mean diameter of the nanowires as reported in their work [5] and a $Q_{lost}$ value of 0.064 C cm$^{-2}$ from this work, the expected first-cycle capacity loss is 1488 mAh g$^{-1}$, which is not too different from the reported value. Considering that the nanotube diameter in their work had a statistical distribution and the difference in electrolyte additives between their work and this work, the agreement can be considered reasonable. The work of Chan *et al.* [5] and other similar investigations [6,22,25] demonstrated that nano-scale Si electrode structures improve mechanical integrity and fracture resistance, which is beneficial for cycle-life. At the same time, these structures also offer large surface area for SEI formation and have higher first cycle loss, which results in lower overall available energy density for subsequent cycles. The experimental data presented here provides a useful practical tool for battery developers in estimating the expected capacity loss due to first cycle SEI-layer formation and in choosing an appropriate particle size distribution that balances mechanical integrity and the first cycle capacity loss.

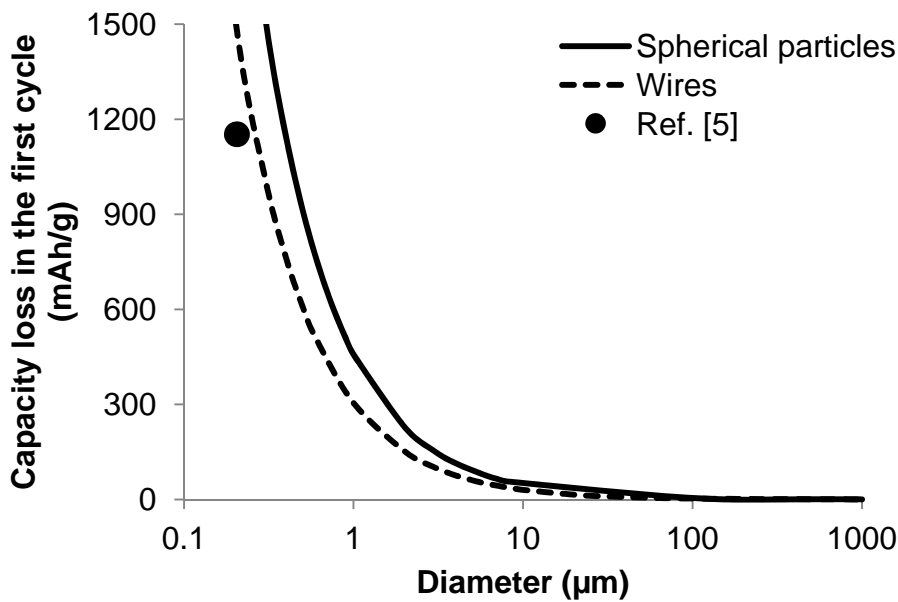

*Figure 8: Estimated capacity loss based on $Q_{\_lost} = 0.064$ C cm$^{-2}$ for different electrode configurations as a function of size (diameter) of Si particles or wires. The symbol represents the data from Ref [5] assuming a nano-wire diameter of 205 nm.*

**3.2 X-ray photo electron spectroscopy (XPS) analysis of Si electrodes**

The elemental surface concentrations of the cycled Si electrodes after different hold times at 0.2 V *vs.* Li/Li$^+$ are shown in Table 1. The carbon concentration does not appear to change during the initial 4 hr and fluctuation in F and O elemental concentrations are observed. The P concentration showed consistent increase during the 10 hr hold time. The changes in elemental concentration suggest changes in the structure of the SEI on Si, which are likely due to thermal reactions of the initial reduction products of the electrolyte.

*Table 1 Elemental concentration on Si negative-electrode surface.*

| 0.2V hold time (hr) | CARBON (%) | FLUORINE (%) | OXYGEN (%) | PHOSPHORUS (%) |
|---|---|---|---|---|
| 0 | 55.2 | 13.6 | 30.7 | 0.5 |
| 4 | 55.2 | 6.3 | 38.1 | 0.4 |
| 6 | 42.6 | 10.5 | 45.5 | 1.4 |
| 10 | 50.2 | 25.6 | 21.7 | 2.2 |

Figure 9 shows that the C1s spectra of the sample charged to 0.2 V *vs.* Li/Li$^+$ contains peaks attributed to the C-O bond of ethers and carbonates at 286-287 eV and the C=O peak from carbonates at 289-290 eV. The carbonate peak increases during the first 6 hr of holding, but decreases in intensity after 10 hr holding.

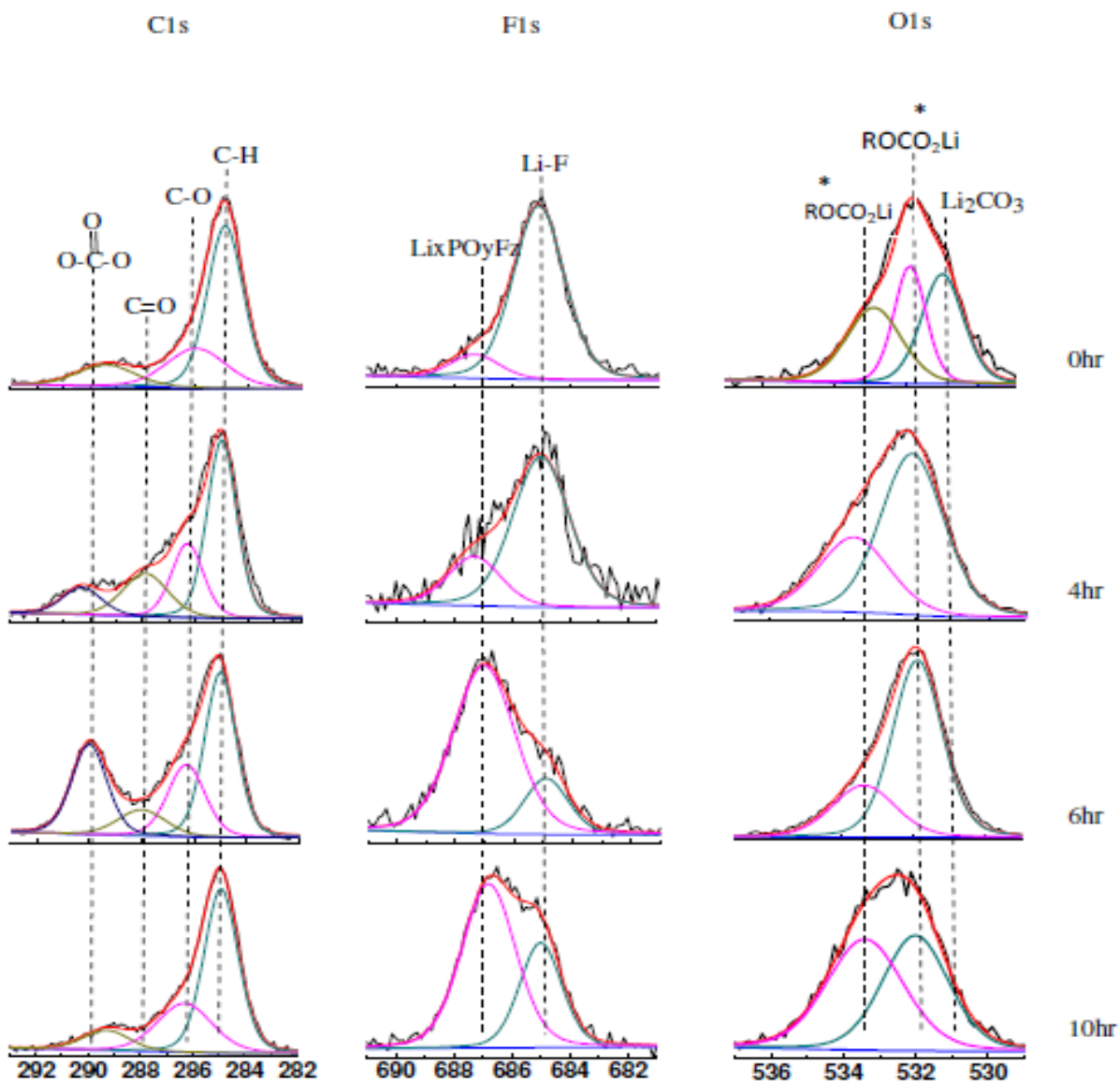

*Figure 9: XPS spectra of Si Negative-electrodes after one complete cycle of lithiation and delithiation.*

The F1s XPS spectra of the sample after initial charging contains a large peak for LiF (685 eV) and a small peak for Li$_x$PO$_y$F$_z$ at 687 eV. As the electrodes are maintained at 0.2 V *vs.* Li/Li$^+$, a steady increase is observed in the concentration of Li$_x$PO$_y$F$_z$. The related P2p peak of Li$_x$PO$_y$F$_z$ also has a similar steady increase in intensity. The O1s spectra contain a broad peak between 531 and 534 eV consistent with the presence of C-O and C=O bonds supporting the presence of ethers and carbonates as suggested by the C1s spectra. Fluctuation in intensities of different O1s peaks are observed, which are cosistent with the changes discussed in the C1s, F1s, and P2p spectra. The XPS data suggests that the initial electrolyte decomposition products on the surface of the electrodes are composed of lithium alkyl carbonates and LiF. However, upon holding the electrode at 0.2 V *vs.* Li/Li$^+$ the composition shifts to higher concentrations of Li$_x$PO$_y$F$_z$ which is likely due to the thermal instability of the LiPF$_6$ salt and its reactivity toward basic species such as lithium alkyl carbonates [26].

Most of the reduction reactions of the electrolyte on the anode surface occur prior to 0.3 V *vs.* Li/Li$^+$. The surface film generated effectively passivates the lithiated Si anode and prevents further electrolyte reduction. However, the initially formed anode SEI is not thermally stable in the presence of LiPF$_6$ electrolytes and changes over the course of storage at low potential. The thermal decomposition reactions of the SEI appear to be limited to the outer layers of the SEI under these conditions since additional current flow and electrolyte reduction does not occur.

## 4. Conclusions

Coin cells were prepared with Si thin film as working and lithium metal foil as counter/reference electrode and 1M of LiPF$_6$ dissolved in EC:DEC:DMC (1:1:1) as electrolyte, and the cells were cycled (*i.e.*, lithiation and delithiation) under different electrochemical conditions to estimate the capacity lost $Q_{lost}$ per unit area of Si electrode due to the formation of SEI layer. Experimental data showed that the $Q_{lost}$ is in the range of 0.06 – 0.1 C cm$^{-2}$ (0.0178 and 0.0278 mAh cm$^{-2}$), which can be used to predict or estimate the first cycle capacity loss for Si electrodes in any configuration for which the real surface area is known. Predictions based on the above $Q_{lost}$ value agreed reasonably well with the data from literature on nano-structured Si electrodes.

To identify the chemical species that consist of the SEI on Si electrodes, XPS analysis was performed. SEI formed on silicon electrodes show species similar to lithium alkyl carbonates, LiF and Li$_x$PO$_y$F$_z$. Lithium alkyl carbonates increase in concentration as the hold time increases up to 6 hrs. While Li$_x$PO$_y$F$_z$ concentration also increases with the hold time. This indicates that SEI components are not stable and further electrolyte degradation occurs thermally.

The experimental data and the parameters presented here provide a useful practical tool for battery developers in estimating the expected capacity loss due to first cycle SEI-layer formation and in choosing an appropriate particle size distribution that balances mechanical integrity and the first cycle capacity loss.

## 5. Acknowledgements

The authors gratefully acknowledge funding from Department of Energy Office of Basic Energy Sciences  EPSCoR Implementation award (DE-SC0007074), and NASA EPSCoR award # NNX10AN03A (MJC). The authors thank Dr. Robert Kostecki at Lawrence Berkeley National Laboratory for help with obtaining Raman spectra on Si (100) and sputtered Si thin films.

# 6. References

1. U. Kasavajjula, C. Wang, A.J. Appleby, J. Power Sources 163 (2007) 1003.

2. S.D. Beattie, D. Larcher, M. Morcrette, B. Simon, J-M. Tarascon, J. Electrochem. Soc. 155 (2008) A158.

3. L.Y. Beaulieu, K.W. Eberman, R.L. Turner, L.J. Krause, J.R. Dahn, Electrochem. Solid-State Lett. 4 (2001) A137.

4. S. Bourderau, T. Brouse, D.M. Schleich, J. Power Sources 81-82 (1999) 233.

5. C.K. Chan, H. Peng, G. Liu, K. McIlwrath, X.F. Zhang, R.A. Huggins, A.Y. Cui, Nature Nanotech. 3 (2008) 31.

6. J. Graetz, C.C. Ahn, R. Yazami, B. Fultz, Electrochem. Solid-State Lett. 6 (2003) A194.

7. R. Fong, U von Sacken, J.R. Dahn, J. Electrochem. Soc. 137 (1990) 2009.

8. D. Aurbach, M. Moshkovich, Y. Cohen, A. Schechter, Langmuir 15 (1999) 2947.

9. D. Aurbach, J. Power Sources 89 (2006) 206.

10. C.C. Nguyen, S. Song, Electrochem. Comm. 12 (2010) 1593.

11. E.J. Peled, Electrochem. Soc. 126 (1979) 2047.

12. Y.M. Lee, J.Y. Lee, H. Shim, J.K. Lee, J. Park, J. Electrochem. Soc.154 (2007) A515.

13. L. Chen, K. Wang, X. Xie, J. Xie, J. Power Sources 174 (2007) 538.

14. S. Dalavi, P.R. Guduru, B.L. Lucht, J. Electrochem. Soc. 159 (2012) A642.

15. V. Etacheri, O. Haik, Y. Goffer, G.A. Roberts, I.C. Stefan, R. Fasching, D. Aurbach, Langmuir 28 (2012) 965.

16. H. Nakai, T. Kubota, A. Kita, A. Kawashima, J. Electrochem. Soc. 158 (2011) A798.

17. V.A. Sethuraman, K. Kowolik, V. Srinivasan, J. Power Sources 196 (2011) 393.

18. Y. Yen, S. Chao, H. Wu, N. Wu, J. Electrochem. Soc. 156 (2009) A95.

19. V.A. Sethuraman, M.J. Chon, M. Shimshak, V. Srinivasan, P.R. Guduru, J. Power Sources 195 (2010) 5062.

20. V.A. Sethuraman, V. Srinivasan, A.F. Bower, P.R. Guduru, J. Electrochem. Soc. 157 (2010) A1253.

21. V.A. Sethuraman, M.J. Chon, M. Shimshak, N. Van Winkle, P.R. Guduru, Electrochem. Comm. 12 (2010) 1614.

22. K. Xu, Chem. Rev. 104 (2004) 4303.

23. M.J. Chon, V.A. Sethuraman, A. McCormick, V. Srinivasan, P.R. Guduru, Phys. Rev. Lett. 107 (2011) 045503.

24. D. Mazouzi, B. Lestriez, L. Roue, D. Guyomard, Electrochem. Solid-State Lett. 12 (2009) A215.

25. J.R. Szczech, S. Jin, Energy Environ. Sci. 4 (2011) 56.

26. W. Li, A. Xiao, B.L. Lucht, M.C. Smart, B.V. Ratnakumar, J. Electrochem. Soc. 155 (2008) A648.